\documentclass[12pt]{article}
\usepackage{epsf}
\newcommand{\eqb}{\begin{equation}}
\newcommand{\eqe}{\end{equation}}
\newcommand{\dmb}{\begin{displaymath}}
\newcommand{\dme}{\end{displaymath}}

\newcommand{\eab}{\begin{eqnarray}}
\newcommand{\eae}{\end{eqnarray}}

\newcommand{\be}{\begin{equation}}
\newcommand{\ee}{\end{equation}}

\def\lsim{\mathrel{\raise.3ex\hbox{$<$\kern-.75em\lower1ex\hbox{$\sim$}}}}
\def\gsim{\mathrel{\raise.3ex\hbox{$>$\kern-.75em\lower1ex\hbox{$\sim$}}}}
\def\Li2{{\rm Li}_2}

\newcommand{\bmG}{\mathbf G}

\newcommand{\msbar}{$\overline{\mbox{MS}}$ }

\begin{document}

\begin{titlepage}
\begin{flushright}
MPI-PhT 2002-86 \\
hep-ph/0212297\\
December 2002
\end{flushright}
\vspace{0.6cm}

\begin{center}
\Large{{\bf Uncertainties in the \msbar Bottom Quark Mass\\
from Relativistic Sum Rules}}

\vspace{1cm}

 G.~Corcella and A.~H.~Hoang 

\end{center}
\vspace{0.3cm}

\begin{center}
{\em Max-Planck-Institut f\"ur Physik\\ 
Werner-Heisenberg-Institut\\ 
F\"ohringer Ring 6, 80805 M\"unchen\\ 
Germany}
\end{center}
\vspace{0.5cm}

\begin{abstract}
\noindent
A detailed compilation of uncertainties in the \msbar  bottom quark mass
$\overline m_b(\overline m_b)$ obtained from low-$n$ spectral sum rules
at order $\alpha_s^2$ is given including charm mass effects and
secondary $b\bar b$ production. The experimental continuum region
above $11.1$~GeV is treated conservatively.
An inconsistency of the PDG averages for the electronic partial
widths of $\Upsilon(\mbox{4S})$ and  $\Upsilon(\mbox{5S})$
is pointed out. From our analysis we obtain $\overline m_b(\overline
m_b)=4.20\pm 0.09$~GeV. The impact of future CLEO data is discussed.

\end{abstract} 

\end{titlepage}

\section*{Introduction}
\label{sectionintroduction}

Present data from B factories on inclusive decays already require
precise knowledge of the bottom quark mass parameter with a numerical
precision of the order $50$~MeV with a reliable estimate of the
uncertainty. This will become even more acute 
in the future when more data becomes available. Using methods
based on perturbative QCD, there have been several approaches in the past
aiming at uncertainties of less than $100$~MeV. The most frequently
used  method is based on large-$n$ ($n\gsim 4$) moments of the $b\bar b$
production cross section in $e^+e^-$ annihilation~\cite{Voloshin1},
\begin{equation}
\label{momdef}
P_n \, = \, \int\frac{ds}{s^{n+1}}\,R_{b\bar b}(s)
\,,
\end{equation}
where $R_{b\bar b}=\sigma(e^+e^-\to b\bar b+X)/\sigma(e^+e^-\to
\mu^+\mu^-)$ and the contributions from the virtual Z are
neglected. From fitting moments obtained from experimental data to
the corresponding theoretical expressions the bottom mass can be
determined in threshold schemes (for recent reviews see
Refs.\,\cite{reviews}).\footnote{
Recently, the bottom mass has been determined in different threshold
schemes from moments of inclusive semileptonic and radiative B meson
decay spectra with an uncertainty of about $100$~MeV, which is
dominated by experimental errors.\cite{Bauer1} 
(See also Ref.\,\cite{Battaglia1}.)
} 
Large-$n$ (``non-relativistic'') moments have
the advantage that the badly known experimental continuum $b\bar b$ cross
section above the $\Upsilon(\mbox{6S})$ is strongly suppressed in
comparison to the rather well known resonance region.
However, the order $\alpha_s^2$ corrections in the framework of the
non-relativistic expansion are generally quite large and small
uncertainties below $100$~MeV can only be achieved with additional
assumptions on higher order corrections.~\cite{reviews} This indicates
that an improvement in the treatment of the non-relativistic bottom
quark dynamics might be needed.   

A different method
uses low-$n$ (``relativistic'') moments~\cite{Novikov1,Reinders1}
where $n\lsim 4$. In the recent past, relativistic moments have been
used less frequently because the badly known continuum region
represents a major source of uncertainty that is not reducible without
additional assumptions. Theoretically, the usual loop expansion in powers
of $\alpha_s$ can be employed since for small $n$ the bottom dynamics
is relativistic. Here, the \msbar mass is an appropriate
mass definition to be used and extracted. In contrast to large-$n$
moments, the low-$n$ moments show a quite good perturbative behavior. 
A recent analysis by K\"uhn and Steinhauser~\cite{Kuhn1} used moments at
order $\alpha_s^2$. Adopting a theory-driven perspective, the
experimental continuum data above the 
$\Upsilon(\mbox{6S})$ was obtained from theoretical results for
$R_{b\bar b}$, basically eliminating uncertainties from the continuum
region. In Ref.\,\cite{Kuhn2} the important conclusion was drawn that,
using the strategy of Ref.\,\cite{Kuhn1}, a substantially more
accurate measurement of the $\Upsilon(\mbox{1S})$ -
$\Upsilon(\mbox{6S})$ region at CLEO~\cite{Shipsey1} 
could result in an uncertainty in $\overline m_b(\overline m_b)$ of
only $30$~MeV.

It is the main purpose of this paper to give a detailed compilation of
all sources of uncertainties in the bottom \msbar mass $\overline
m_b(\overline m_b)$ obtained from low-$n$ moments, including
a more conservative treatment of the experimental continuum region. 
We believe that this compilation can contribute to a more
differentiated view on the current uncertainties in 
$\overline m_b(\overline m_b)$ from low-$n$ moments and on the impact
of new more precise data in the $\Upsilon$ resonance region from
CLEO. In our analysis we also include the contributions from secondary
$b\bar b$ 
production from gluon splitting and the effects of the non-zero charm
quark mass, which have to our knowledge not been taken into account
before. Both effects turn out to be small. Finally, we point out an
inconsistency in the way the PDG has treated the original results for
the electronic partial widths of $\Upsilon(\mbox{4S})$ and
$\Upsilon(\mbox{5S})$ from CUSB~\cite{CUSB} and CLEO~\cite{CLEO},
which leads to a contribution to the 
experimental moments $P_n$ that is smaller than the contributions
obtained from the original data in that energy region, both from CUSB
and CLEO.

\section*{Theoretical Moments}
\label{sectiontheory}

For the QCD parameters used in this work we adopt the \msbar
renormalization scheme and the convention that the bottom quark
participates in the running ($n_f=5$). The masses of the quarks in the
first two generations are set to zero. In terms of $\overline
m_b(\mu)$ the moments in the OPE, including the known perturbative
corrections to order $\alpha_s^2$ and the contribution from the
dimension four gluon condensate, take the form
\begin{eqnarray}
P_n 
& = &
\frac{1}{(\,4\overline m_b(\mu)\,)^{n}}\,\bigg\{\,
f_n^0 \, + \, 
\bigg(\frac{\alpha_s(\mu)}{\pi}\bigg)\,
  \bigg(\, f_n^{10} \, + \, 
         f_n^{11}\,\ln\Big(\frac{\overline m_b^2(\mu)}{\mu^2}\Big)
  \,\bigg)
\nonumber\\[2mm]
& &  \, + \, 
\bigg(\frac{\alpha_s(\mu)}{\pi}\bigg)^2\,
  \bigg(\, f_n^{20}(r) \, + \, 
         f_n^{21}\,\ln\Big(\frac{\overline m_b^2(\mu)}{\mu^2}\Big)\, + \, 
         f_n^{22}\,\ln^2\Big(\frac{\overline m_b^2(\mu)}{\mu^2}\Big)
  \,\bigg)
\nonumber\\[2mm]
& & \, + \,
\frac{\langle{\textstyle\frac{\alpha_s}{\pi}}\bmG^2\rangle}
{(\,4\overline m_b(\mu)\,)^2}\,
  \bigg[\,g_n^{0} + 
\bigg(\frac{\alpha_s(\mu)}{\pi}\bigg)\,
  \bigg(\, g_n^{10} \, + \, 
         g_n^{11}\,\ln\Big(\frac{\overline m_b^2(\mu)}{\mu^2}\Big)
  \,\bigg)
  \,\bigg]
\,\bigg\}
\,,
\label{Mn1}
\end{eqnarray} 
where $r\equiv m_c/m_b$. In terms of the more specific choice of $\overline
m_b(\overline m_b)$, the moments have the simpler form
\begin{eqnarray}
P_n 
& = &
\frac{1}{(\,4\overline m_b(\overline m_b)\,)^{n}}\,\bigg\{\,
f_n^0 \, + \, 
\bigg(\frac{\alpha_s(\mu)}{\pi}\bigg)\,f_n^{10} 
\nonumber\\[2mm]
& &  \, + \, 
\bigg(\frac{\alpha_s(\mu)}{\pi}\bigg)^2\,
  \bigg(\, f_n^{20}(r) \, - \, 
         \frac{1}{4}\,\beta_0\,f_n^{10}\,
   \ln\Big(\frac{\overline m_b^2(\overline m_b)}{\mu^2}\Big)
  \,\bigg)
\nonumber\\[2mm]
& & \, + \,
\frac{\langle{\textstyle\frac{\alpha_s}{\pi}}\bmG^2\rangle}
{(\,4\overline m_b(\overline m_b)\,)^2}\,
  \bigg[\,g_n^{0} + 
\bigg(\frac{\alpha_s(\mu)}{\pi}\bigg)\, g_n^{10} 
  \,\bigg]
\,\bigg\}
\,,
\label{Mn2}
\end{eqnarray} 
where $\beta_0=11-2/3 n_f$.
The Born and order $\alpha_s$ terms of the moments are known
since a long time~\cite{Novikov1,Reinders1} and the order $\alpha_s^2$
contributions for primary $b\bar b$ production for massless light
quarks have been determined in Ref.\,\cite{Chetyrkin1}. We have
cross-checked these contributions with the explicit expressions for
the corresponding contributions to $R_{b\bar b}$ given in
Ref.\,\cite{Chetyrkin2}. The order $\alpha_s^2$ contributions to the moments
from secondary $b\bar b$ production, where the $b\bar b$ pair is
produced through gluon radiation off light quarks, has been computed
from the corresponding results for the $R$-ratio given in
Refs.\,\cite{Hoang1,Hoang2}. These contributions only affect the
coefficient $f_n^{20}$. The coefficients of the gluon condensate
have been taken from Ref.\,\cite{Broadhurst1,Baikov1}.
For convenience, the numerical results for the coefficients in
Eqs.~(\ref{Mn1}) and (\ref{Mn2}) for $n=1,2,3,4$ and $m_c=0$ ($r~=~0$)
are collected in Tab.\,\ref{tabcoeff}.
\begin{table}[t!]  
\begin{center}
\begin{tabular}{|c||r|r|r|r|} \hline
$n$ & $1$\mbox{\hspace{6mm}} & $2$\mbox{\hspace{6mm}} & 
      $3$\mbox{\hspace{6mm}} & $4$\mbox{\hspace{6mm}} \\ \hline\hline 
$f_n^{0}$ & $ 0.2667$ & $ 0.1143$ & $ 0.0677$ & $ 0.0462$ \\ \hline 
 $f_n^{10}$ & $ 0.6387$ & $ 0.2774$ & $ 0.1298$ & $ 0.0508$ \\ \hline 
 $f_n^{11}$ & $ 0.5333$ & $ 0.4571$ & $ 0.4063$ & $ 0.3694$ \\ \hline 
 $f_n^{20}(0)$ & $ 0.9446$ & $ 0.8113$ & $ 0.5172$ & $ 0.3052$ \\ \hline 
 $f_n^{21}$ & $ 0.8606$ & $ 1.2700$ & $ 1.1450$ & $ 0.8682$ \\ \hline 
 $f_n^{22}$ & $ 0.0222$ & $ 0.4762$ & $ 0.8296$ & $ 1.1240$ \\ \hline 
 $g_n^{0}$ & $ -4.011$ & $ -6.684$ & $ -9.722$ & $-13.088$ \\ \hline 
 $g_n^{10}$ & $ -4.876$ & $  1.386$ & $ 16.964$ & $ 44.081$ \\ \hline 
 $g_n^{11}$ & $ -24.063$ & $ -53.473$ & $ -97.224$ & $-157.055$ \\ \hline
\end{tabular}
\caption{\label{tabcoeff} 
Coefficients of the theoretical expressions for the moments $P_n$ to
order $\alpha_s^2$ for massless light quarks. 
}
\end{center}
\vskip 3mm
\end{table}
The numbers for the $f_n$'s agree with Ref.\,\cite{Kuhn1} up to a
convention dependent factor of $1/4$, except for the results for 
$f_n^{20}(0)$, which are slightly larger accounting for the
contributions from secondary  $b\bar b$ production. The effects of the
non-zero charm quark mass are 
generated either through virtual gluon self-energy effects or through
real primary or secondary associated charm production. The corresponding
contributions to $R_{b\bar b}$ for arbitrary mass constellations have
been given in Refs.\,\cite{Chetyrkin2,Hoang1}. Numerical values of the
charm quark mass corrections to the coefficient $f_n^{20}$ are
displayed in Tab.\,\ref{tabcharm} for values of $r$ between $0.1$ and
$0.5$. 
\begin{table}[t!]  
\begin{center}
\begin{tabular}{|c||c|c|c|c|c|} \hline
$r$ & $0.1$ & $0.2$ & $0.3$ & $0.4$ & $0.5$\\ \hline\hline
 $f_1^{20}(r)-f_1^{20}(0)$ & $-0.0021$ & $-0.0078$ & $-0.0164$ &
 $-0.0266$ & $-0.0382$ 
\\ \hline 
 $f_2^{20}(r)-f_2^{20}(0)$ & $-0.0028$ & $-0.0091$ & $-0.0187$ &
 $-0.0302$ & $-0.0430$ 
\\ \hline 
 $f_3^{20}(r)-f_3^{20}(0)$ & $-0.0024$ & $-0.0101$ & $-0.0204$ &
 $-0.0330$ & $-0.0466$ 
\\ \hline 
 $f_4^{20}(r)-f_4^{20}(0)$ & $-0.0030$ & $-0.0109$ & $-0.0219$ &
 $-0.0348$ & $-0.0491$ 
\\ \hline
\end{tabular}
\caption{\label{tabcharm} 
Corrections due to the non-zero charm quark mass to the order
$\alpha_s^2$ coefficient $f_n^{20}$ for $r=m_c/m_b$.
}
\end{center}
\vskip 3mm
\end{table}
We note that the numbers given in Tab.\,\ref{tabcharm} also include
the non-zero charm mass effects in the bottom quark pole-\msbar mass
relation~\cite{Broadhurst1}.

\section*{Experimental Moments}
\label{sectionexperiment}

\begin{table}[t!]  
\begin{center}
\begin{tabular}{|c||c|c|c|c|} \hline
 & $P_1$ & $P_2$ & $P_3$ & $P_4$  \\
\raisebox{1.5ex}[-1.5ex]{contribution} & 
 $\times\,10^3~\mbox{GeV}^2$ & $\times\,10^5~\mbox{GeV}^4$ & 
 $\times\,10^7~\mbox{GeV}^6$ & $\times\,10^9~\mbox{GeV}^8$ \\
\hline\hline 
$\Upsilon(\mbox{1S})$ & $ 0.766(29)$ & $ 0.856(32)$ & $ 0.956(36)$ & $ 1.068(40)$ 
\\ \hline 
$\Upsilon(\mbox{2S})$ & $ 0.254(16)$ & $ 0.252(16)$ & $ 0.251(15)$ & $ 0.250(15)$ 
\\ \hline 
$\Upsilon(\mbox{3S})$ & $ 0.211(29)$ & $ 0.196(27)$ & $ 0.183(26)$ & $ 0.171(24)$ 
\\ \hline 
$[\Upsilon(\mbox{4S})-\Upsilon(\mbox{5S})]_{\rm PDG}$ & $ 0.222(40)$ & $ 0.192(34)$ & $ 0.167(29)$ & $ 0.145(25)$ 
\\ \hline 
$[\Upsilon(\mbox{4S})-\Upsilon(\mbox{5S})]_{\rm CUSB}$ & $ 0.257(42)$ & $ 0.223(36)$ & $ 0.194(31)$ & $ 0.169(27)$ 
\\ \hline 
$[\Upsilon(\mbox{4S})-\Upsilon(\mbox{5S})]_{\rm CLEO}$ & $ 0.244(95)$ & $ 0.213(82)$ & $ 0.186(72)$ & $ 0.162(62)$ 
\\ \hline 
$[\Upsilon(\mbox{4S})-\Upsilon(\mbox{5S})]_{\rm our}$ & $ 0.251(95)$ & $ 0.218(82)$ & $ 0.190(72)$ & $ 0.165(62)$ 
\\ \hline 
$\Upsilon(\mbox{6S})$ & $ 0.048(11)$ & $ 0.039(9)$ & $ 0.032(7)$ & $ 0.027(6)$ 
\\ \hline 
$11.1~\mbox{GeV}-12.0~\mbox{GeV}$ & $ 0.418(57)$ & $ 0.314(44)$ & $ 0.236(34)$ & $ 0.178(27)$ 
\\ \hline 
$12.0~\mbox{GeV}-M_Z$ & $ 2.467(26)$ & $ 0.886(21)$ & $ 0.414(13)$ & $ 0.217(8)$ 
\\ \hline 
$M_Z-\infty$ & $ 0.047(1)$ & $ 0.000(0)$ & $0.000(0)$ & $0.000(0)$ 
\\ \hline
\end{tabular}
\caption{\label{tabdatamom} 
Individual contributions to the experimental moments including uncertainties.
The contribution from a resonance $k$ has been determined in the narrow width
approximation, 
$(P_n)_k=9\pi\,\Gamma^{e^+e^-}_k/[\alpha(10\,\mbox{GeV})\,M_k^{2n+1}]$,
where for the electromagnetic coupling
$[\alpha(10~\mbox{GeV})]^{-1}=131.8$ has been adopted.
}
\end{center}
\vskip 3mm
\end{table}
For the contributions to the experimental moments from the
$\Upsilon(\mbox{1S})$, $\Upsilon(\mbox{2S})$, $\Upsilon(\mbox{3S})$
and $\Upsilon(\mbox{6S})$ we use the averages for masses and $e^+e^-$
widths given by the PDG~\cite{PDG}. In Tab.\,\ref{tabdatamom} a
collection of all contributions to the moments including
uncertainties is given. The averages for the $\Upsilon(\mbox{1S})$,
$\Upsilon(\mbox{2S})$ and $\Upsilon(\mbox{3S})$ are dominated by data
from ARGUS~\cite{ARGUS} and the averages for the
$\Upsilon(\mbox{6S})$ are from results from CUSB~\cite{CUSB} and
CLEO~\cite{CLEO}.  

In the 4S--5S region between $10.5$ and $10.95$~GeV there have been
measurements from  CUSB~\cite{CUSB} and CLEO~\cite{CLEO}. We find it
remarkable that both experiments observed an additional resonance-like
enhancement between the $\Upsilon(\mbox{4S})$ and $\Upsilon(\mbox{5S})$
at about $10.7$~GeV. Whereas CUSB fitted for $\Upsilon(\mbox{4S})$ and
$\Upsilon(\mbox{5S})$ resonances within in a coupled channel model,
the CLEO experiment was fitting an additional resonance, called
``$B^*$'', at $m_{B^*}=10.684\pm 0.013$~GeV with an $e^+e^-$ width of
$\Gamma^{e^+e^-}_{B^*}= 0.20\pm 0.11$~keV. As a consequence the
$e^+e^-$ widths for $\Upsilon(\mbox{4S})$ and $\Upsilon(\mbox{5S})$
obtained from CLEO were systematically smaller than those from
CUSB. In the PDG compilation, however, the existence of the
``$B^*$'' contribution in the CLEO analysis was ignored when the
averages for $\Upsilon(\mbox{4S})$ and 
$\Upsilon(\mbox{5S})$ have been determined.\footnote{
The PDG number for the electronic width of the $\Upsilon(\mbox{4S})$ 
is a weighted average of the results from CUSB~\cite{CUSB},
CLEO~\cite{CLEO} and ARGUS~\cite{ARGUS}, where the ARGUS result agrees
better with the one from CUSB. 
} As a consequence, the
contribution to the moments from the region between $10.5$ and
$10.95$~GeV using the PDG averages is, although compatible within
errors, systematically lower than the
contributions one obtains using the numbers given in the original
CLEO and CUSB publications (see Tab.\,\ref{tabdatamom}). For our
analysis we decided to ignore the PDG averages and to take the
average of the original CLEO (including the ``$B^*$'') and the CUSB
contributions to the moments for this energy region.
We adopted the larger CLEO error assuming that the 
respective uncertainties are correlated. We believe that this
conservative treatment is justified as long as the situation is not
clarified. We note that the PDG
treatment of the 4S--5S region does not affect the results for the
hadronic vacuum polarization effects of $\alpha(M_Z)$ and $(g-2)_\mu$
because the corresponding differences are much smaller than the total
uncertainties.  
For the $\Upsilon(\mbox{6S})$ we used the PDG averages
since it is unlikely that the different treatment of
the enhancement observed between $\Upsilon(\mbox{4S})$ and
$\Upsilon(\mbox{5S})$ has affected the fits above the
$\Upsilon(\mbox{5S})$.  

There is no direct experimental data for $\sigma(e^+e^-\to \gamma^*
\to b\bar b+X)$ in the region above $11.1$~GeV. However, there are
measurements of the total hadronic cross section taken by a number of
experiments up to energies close to $M_Z$ that are compatible with
the Standard Model predictions. Furthermore, from
measurements of $R_b$ at the Z pole by LEP and in the region
between $133$ and $207$~GeV by LEP2\footnote{
At the Z pole $R_b$ is defined as the ratio of the total b quark
partial width of the Z to its total hadronic partial width,
$\Gamma_{b\bar b}/\Gamma_{\rm had}$, and 
for LEP2 energies it is defined as the ratio of the total $b\bar b$
cross section to the total hadronic one.
}, it is known that perturbative QCD
agrees with the data for $b\bar b$ production to about $1\%$ at $M_Z$
and to about $10\%$ in the LEP2 region~\cite{LEP2}. It is therefore
not unreasonable to estimate the   
experimental contribution to the moments from above the
$\Upsilon(\mbox{6S})$ 
from perturbation theory itself. Although order $\alpha_s^3$
corrections to $R_{b\bar b}$ in the high energy expansion are known,
we use the perturbative $b\bar b$ cross section to order $\alpha_s^2$
to estimate the continuum contributions because the theoretical
moments are likewise only available to order $\alpha_s^2$. A much more
subtle question is how to estimate the ``experimental'' uncertainties
in this region. The approach of Ref.\,\cite{Kuhn1} assumes that the
experimental  
data for the $b\bar b$ cross section including errors lie within the
theoretical predictions and uses the small theoretical
errors. This approach is quite similar to using
finite energy sum  rules~\cite{Schilcher1} where an upper ``duality''  
cutoff $s_{\rm max}\gsim 11.1$~GeV is used 
in the integral in Eq.\,(\ref{momdef}).

In Tab.\,\ref{tabdatamom} we have displayed the continuum
contributions to the moments to order $\alpha_s^2$.
The charm quark mass has been set to zero. We have subdivided the
continuum contribution into three parts coming from 
$11.1-12.0$~GeV (region~1), $12~\mbox{GeV}-M_Z$ (region~2) and 
$M_Z-\infty$ (region~3) in order to visualize the impact of the
various energy regions. The theoretical errors shown in
Tab.\,\ref{tabdatamom} come from varying the strong coupling in the
range $\alpha_s(M_Z)=0.118\pm 0.003$ and the \msbar bottom mass in the
conservative PDG range $\overline m_b(\overline m_b)=4.2\pm 0.2$.
The renormalization scale $\mu$ has been varied between $2.5$ and
$10$~GeV. For the running of $\alpha_s$ and 
$\overline m_b$ four-loop renormalization group equations
have been used.  Our theoretical errors are larger than 
in Ref.\,\cite{Kuhn1} where also order $\alpha_s^3$ contributions have
been included~\cite{Steinhauser1}. Our central values
have been obtained from the average of the respective extremal
values. Region~1 has been displayed separately because data for
the $b\bar b$ cross section could potentially be collected there by 
CLEO~\cite{Shipsey1} in the near future. 

Since the continuum region above $11.1$~GeV is unsuppressed and
constitutes a sizeable contribution to the experimental
moments\footnote{ 
We note that the relative contribution to the low-$n$
$c\bar c$ spectral moments coming from continuum energies above
$4.5$~GeV is considerably smaller than for the continuum region above 
$11.1$~GeV in $b\bar b$ spectral sum rules, see e.\,g.\,Ref.\,\cite{Kuhn1}.  
},
using the small theory errors leads to a considerable model-dependence
of the bottom quark mass.
For our final error estimate of $\overline
m_b(\overline m_b)$ we adopt a $10\%$ correlated (relative) error for
the continuum regions~2 and 3 and ignore the respective theory errors
given in Tab.\,\ref{tabdatamom}.  This choice is, in principle, as
arbitrary as using the theoretical errors (or no errors at all), but
should reduce the model-dependence to an acceptable level. For
region~1 we use 
the theory error since here the variation of the \msbar mass in the 
conservative PDG bounds has the largest impact and leads to a
variation of more than $10\%$.

\section*{Uncertainties in $\overline m_b(\overline m_b)$}
\label{sectionmm}

For the determination of $\overline m_b(\overline m_b)$ and the
uncertainties we have used 4 methods:
\begin{enumerate}
\item The bottom mass $\overline m_b(\overline m_b)$ is determined
  from single moment fits ($n=1,2,3$) using Eq.\,(\ref{Mn2}).
\item The bottom mass $\overline m_b(\mu)$ is determined from single
  moment fits ($n=1,2,3$) using Eq.\,(\ref{Mn1}) and  $\overline
  m_b(\overline m_b)$ is computed subsequently using renormalization
  group equations.
\item The bottom mass $\overline m_b(\overline m_b)$ is determined
  from fits to ratios $P_n/P_{n+1}$ ($n=1,2$) using Eq.\,(\ref{Mn2}).
\item The bottom mass $\overline m_b(\mu)$ is determined from fits to
  ratios $P_n/P_{n+1}$ ($n=1,2$) using Eq.\,(\ref{Mn1}) and  $\overline
  m_b(\overline m_b)$ is computed subsequently using renormalization
  group equations.
\end{enumerate}
For the analysis we employed only moments for $n=1,2,3$ to avoid the
large higher order contributions $\sim(\alpha_s\sqrt{n})^k$ that are
characteristic for the large-$n$ moments and need to be summed. For
method~3 and 4 we did not expand the perturbative series in the
theoretical ratios $P_n/P_{n+1}$. We checked that expanding the
theoretical ratios has only very small effects on the results.
We employed four-loop renormalization group equations
and used $\alpha_s(M_Z)=0.118\pm 0.003$,
$m_c=1.3\pm 0.2$~GeV, 
$\langle{\textstyle\frac{\alpha_s}{\pi}}\bmG^2\rangle
=(0.024\pm 0.024)~\mbox{GeV}^4$ as theoretical input. The
renormalization scale $\mu$ was varied between $2.5$ and $10$~GeV.

\begin{table}[t!]  
\begin{center}
\begin{tabular}{|c||c|c|c||c|c|} \hline
 & \multicolumn{3}{|c||}{Method 1 (2)} & 
   \multicolumn{2}{|c|}{Method 3 (4)}
\\ \hline
$n$ & 1 & 2 & 3 & 1 & 2 
\\ \hline\hline
central & 4210 (4214) & 4200 (4205) & 4197 (4200) & 4191 (4195) & 4191 (4191)
\\ \hline\hline
$\Upsilon(\mbox{1S})$ & 14 (13) & 12 (12) & 11 (11) & 11 (11) & 9 (9)
\\ \hline
$\Upsilon(\mbox{2S})$ & 7 (7) & 6 (6) & 5 (5) & 4 (4) & 3 (3) 
\\ \hline
$\Upsilon(\mbox{3S})$ & 14 (14) & 10 (10) & 8 (8) & 7 (7) & 3 (3) 
\\ \hline
4S-5S & 45 (44) & 32 (32) & 22 (22) & 18 (18) & 4 (4) 
\\ \hline
$\Upsilon(\mbox{6S})$ & 5 (5) & 3 (3) & 2 (2) & 2 (2) & 0 (0)
\\ \hline\hline
combined & 67 (67) & 50 (50) & 38 (38) & 33 (33) & 15 (15)
\\ \hline\hline
\mbox{}\hspace{-3.5mm}[region 1]$_{\rm th}$\hspace{-1mm} & 
           27$_{\rm th}$ (26$_{\rm th}$) & 17$_{\rm th}$ (17$_{\rm th}$) & 
           11$_{\rm th}$ (11$_{\rm th}$) &
           7$_{\rm th}$ (7$_{\rm th}$) & 2$_{\rm th}$ (2$_{\rm th}$)
\\ \hline
\mbox{}\hspace{-3.5mm}[region 2]$_{\rm th}$\hspace{-1mm} & 
           12$_{\rm th}$ (12$_{\rm th}$) & 8$_{\rm th}$ (8$_{\rm th}$) &
           4$_{\rm th}$ (4$_{\rm th}$) &
           4$_{\rm th}$ (4$_{\rm th}$) & 4$_{\rm th}$ (4$_{\rm th}$)
\\ \hline
\mbox{}\hspace{-1mm}[region 2]$_{10\%}$\hspace{-1mm} & 
           115 (114) & 33 (33) & 13 (13) & 49 (49) & 29 (29)
\\ \hline 
\mbox{}\hspace{-3.5mm}[region 3]$_{\rm th}$\hspace{-1mm}  & 
           1$_{\rm th}$ (1$_{\rm th}$) & 0$_{\rm th}$ (0$_{\rm th}$) & 
           0$_{\rm th}$ (0$_{\rm th}$) & 
           1$_{\rm th}$ (1$_{\rm th}$) & 0$_{\rm th}$ (0$_{\rm th}$)
\\ \hline
\mbox{}\hspace{-1mm}[region 3]$_{10\%}$\hspace{-1mm}
         & 2 (2) & 0 (0) & 0 (0) & 2 (2) & 0 (0)
\\ \hline
$\delta \langle{\textstyle\frac{\alpha_s}{\pi}}\bmG^2\rangle$
         & 0 (0) & 0 (0) & 0 (0) & 0 (1) & 0 (1)
\\ \hline
$\delta m_c$ & 0 (0) & 0 (0) & 0 (0) & 0 (0) & 0 (0)
\\ \hline
$\delta\alpha_s(M_Z)$ & 17 (18) & 10 (11) & 6 (6) & 3 (3) & 2 (2)
\\ \hline
$\delta\mu$ & 23 (5) & 16 (14) & 11 (27) & 15 (27) & 3 (50)
\\ \hline\hline
combined & 184 (166) & 77 (75) & 41 (57) & 76 (88) & 37 (85)
\\ \hline\hline
total & 251 (233) & 127 (125) & 79 (95) & 110 (121) & 51 (99) 
\\ \hline
\end{tabular}
\caption{\label{taberrors} 
Central values and uncertainties for $\overline m_b(\overline m_b)$ in
units of MeV based on the methods described in the text. 
}
\end{center}
\vskip 3mm
\end{table}
In Tab.\,\ref{taberrors} the results of our analysis for $\overline
m_b(\overline m_b)$ are displayed in detail. The table shows the
respective central values (in units MeV), which were obtained using the
experimental and theoretical central values given before and
$\mu=5$~GeV. The central values obtained with the four methods are
within $15$~MeV around $4.20$~GeV. All uncertainties (in units of MeV)
are presented separately. The experimental errors 
correspond to the uncertainties given in Tab.\,\ref{tabdatamom}, where
for the 4S-5S region our conservative CUSB-CLEO average has been used.
The uncertainty from the 4S-5S region constitutes the largest
experimental error from the resonance region.
The errors from the continuum regions indicated with a subscript
"th" are obtained from the corresponding theory errors shown
Tab.\,\ref{tabdatamom}. For the continuum regions~2 and 3 also the
errors coming from a $10\%$ deviation from the theory prediction  are 
displayed (having no subscript). We note that the latter errors
scale roughly linearly, i.e.\,assuming a $5\%$ ($20\%$) deviation the 
error decreases (increases) by a factor of two, etc.. This illustrates
how strongly the bottom quark mass 
depends on assumptions for the experimentally unknown $b\bar b$
continuum cross section above the $\Upsilon(\mbox{6S})$. 
The theoretical errors have been obtained by
varying each of the theoretical parameters in the ranges given above
while the respective other parameters were fixed to their central
values. For method~3 and 4, which are based on fitting ratios
$P_n/P_{n+1}$, the same theoretical input parameters have been used
for the moments in the numerator and those in the denominator.
If $\alpha_s(M_Z)$ is chosen independently for the moments in the
numerator and denominator, the errors for method~3~(4) are
38~(26)~MeV for $n=1$ and 39~(13)~MeV for $n=2$.
If $\mu$ is chosen independently for the moments in the
numerator and denominator, the errors for method~3 are
52~MeV for $n=1$ and 63~MeV for $n=2$.
We found that, except for the variations of $\mu$, all resulting
errors scale linearly with changes of the input parameters. The
errors coming from variations of $\mu$ have been chosen to be the
larger ones of the two deviations obtained in the ranges
$2.5~\mbox{GeV}<\mu<5$~GeV and 
$5~\mbox{GeV}<\mu<10$~GeV. Note that the overall shift in the central
value of $\overline m_b(\overline m_b)$ coming from the gluon
condensate contribution is between $-0.1$ and $-1$~MeV. 
The shift caused by the non-zero charm quark mass is between $-1$ and
$-3$~MeV, which is an order of magnitude smaller than for large-$n$
moments, where the charm mass effects are enhanced by a factor
$1/\alpha_s^2\sim n$~\cite{Hoangcharm}. Using the PDG average for the 4S-5S
region instead of the CUSB-CLEO average (see Tab.\,\ref{tabdatamom}),
$\overline m_b(\overline m_b)$ is shifted by $10$ to $15$~MeV for
method 1 and 2 and by $2$ to $6$~MeV for method 3 and 4.
We consider the numbers given in
Tab.\,\ref{taberrors} as the main result of this work.

In order to obtain combined errors from the uncertainties of the
resonance data we treated one half of each error as correlated (being
added linearly) and the other half uncorrelated (being added
quadratically) because all data came from $e^+e^-$ machines with
common systematic uncertainties and, roughly, systematic and
statistical uncertainties were found to have comparable
sizes~\cite{CLEO}. 
In Tab.\,\ref{taberrors} the resulting combined resonance errors are
shown in the line below the numbers for the $\Upsilon(\mbox{6S})$.
The theoretical uncertainties including the errors adopted for the
three continuum regions have been combined linearly since they do not
have any statistical meaning and, in particular, the division into
three continuum 
regions is completely arbitrary. Note that for regions~2 and 3 the
uncertainties from the $10\%$ variation of the theory prediction
have been adopted and not the smaller theoretical errors. The
resulting combined error is displayed in the line below the 
numbers for $\delta\mu$.  
To obtain our total error (last line in Tab.\,\ref{taberrors}) we
added the resonance, the continuum and the theory errors linearly. 
As expected we find that the uncertainties from the continuum have the
largest impact for the moments with $n=1$ and $2$ and that they are
partially canceled when ratios are used for the fitting. However, the
theoretical errors are larger for fits with ratios than with single
moments, if the theoretical parameters are chosen independently
for numerator and denominator. In general,
the total error decreases for larger $n$.  
This trend does, however, not continue for higher values $n>3$
particularly for methods~2 and 4. Compared to the results of 
Ref.\,\cite{Kuhn1} our total errors are much larger, particularly for  
fits involving $P_1$ and $P_2$. This is mainly because we adopted more
conservative errors for the continuum region in the experimental
moments, and we combined uncertainties linearly, when they cannot be
treated statistically.
Since we believe that $P_3$ can be computed reliably using
Eqs.\,(\ref{Mn1}) and (\ref{Mn2}), we adopt the average of the total
errors in the third column of 
Tab.\,\ref{taberrors} as our final estimate for the uncertainty and
obtain (rounded to units of $10$~MeV)
\begin{equation}
\overline m_b(\overline m_b) \, = \, 4.20 \, \pm \, 0.09~\mbox{GeV}
\,.
\label{finalresult}
\end{equation}
We do not take into account the small $50$~MeV error from 
$P_2/P_3$ for method~3 because the small error from variations of
$\mu$ only persists if the same $\mu$ is chosen for both moments. 
For an independent choice the error is considerably larger (see text
above). If a $20\%$ relative uncertainty is assumed for the continuum
regions~1, 2 and 3, the final error increases by $20$~MeV.
Our result in Eq.\,(\ref{finalresult}) is compatible with the result
from Ref.\,~\cite{Kuhn1}. Our result is also compatible with results
from 
fitting large-$n$ moments at NNLO in the non-relativistic
expansion~\cite{reviews} and with recent results, using different
methods, obtained in Refs.\,\cite{Schilcher1,Erler1,Eidemuller1}.  

Based on the numbers given in Tab.\,\ref{tabdatamom} and
\ref{taberrors} it is straightforward to discuss the impact of
improved measurements of the resonance parameters at
CLEO. Assuming improved measurements for the electronic widths of 
$\Upsilon(\mbox{1S})$, $\Upsilon(\mbox{2S})$ and
$\Upsilon(\mbox{3S})$~\cite{Shipsey1} at the level of $2\%$, the
combined resonance errors shown in Tab.\,\ref{tabdatamom} would be
reduced by a factor $2/3$, which 
would reduce the total error (obtained from the third
column) by about $10$~MeV. An improved
measurement of the 4S-5S region and the $\Upsilon(\mbox{6S})$  with
the same precision would result approximately in a reduction of the
total error by an additional $10$~MeV. 
A further reduction of the
error below about $70$~MeV, however, will be difficult to achieve
without real experimental data for $\sigma(e^+e^-\to b\bar b)$ in the
continuum region above the  $\Upsilon(\mbox{6S})$ with a precision of
better than ten percent.

\section*{Conclusion}
\label{sectionconclusion}

We have given a detailed analysis of the uncertainties in the \msbar
bottom quark mass $\overline m_b(\overline m_b)$ using low-$n$
spectral moments of the cross section $\sigma(e^+e^-\to \gamma^*\to
b\bar b)$. For the experimental moments we employed experimental data
for the $\Upsilon(\mbox{1S})$ - $\Upsilon(\mbox{6S})$ resonances and
the order $\alpha_s^2$ QCD predictions for the continuum region above
$11.1$~GeV. For the 4S--5S region between $10.5$ and $10.95$~GeV we
found that the PDG averages for the electronic partial widths of
$\Upsilon(\mbox{4S})$ and $\Upsilon(\mbox{5S})$ based on data from
CUSB, CLEO and ARGUS contain an inconsistency, stemming from the fact 
CLEO also assumed the existence of an additional resonance at about
$10.7$~GeV which was ignored in the averaging procedure. For our
analysis we therefore used the original CUSB and CLEO results. For the
continuum region above the $\Upsilon(\mbox{6S})$ we assumed a $10\%$
error in our final error estimate. For the theoretical moments we used
perturbative results at order $\alpha_s^2$ including also the
contributions from secondary $b\bar b$ production and finite charm
mass effects. As our final result we get
$\overline m_b(\overline m_b)=4.20\pm \, 0.09$~GeV and we conclude
that more precise data for the electronic partial widths of the
$\Upsilon$ resonances at CLEO could reduce the error by about
$20$~MeV.

\section*{Acknowledgements}
We thank Ed Thorndike for providing details of the analysis in
Ref.\,\cite{CLEO}.

\vspace{1cm}

\bibliographystyle{prsty}

\begin{thebibliography}{10}
%
%
\bibitem{Voloshin1}
M.~B.~Voloshin,
Int.\ J.\ Mod.\ Phys.\ A {\bf 10}, 2865 (1995)
[arXiv:hep-ph/9502224].
%
%
\bibitem{reviews}
A.~H.~Hoang,
arXiv:hep-ph/0204299;\\
%
A.~X.~El-Khadra and M.~Luke,
arXiv:hep-ph/0208114.
%
%
\bibitem{Bauer1}
C.~W.~Bauer, Z.~Ligeti, M.~Luke and A.~V.~Manohar,
arXiv:hep-ph/0210027.
%
%
\bibitem{Battaglia1}
M.~Battaglia {\it et al.},
arXiv:hep-ph/0210319.
%
%
\bibitem{Novikov1}
V.~A.~Novikov {\it et al.}, 
Phys.\ Rev.\ Lett.\  {\bf 38}, 626 (1977)
[Erratum-ibid.\  {\bf 38}, 791 (1977)].
%
%
\bibitem{Reinders1}
L.~J.~Reinders, H.~R.~Rubinstein and S.~Yazaki,
Nucl.\ Phys.\ B {\bf 186}, 109 (1981).
%
%
\bibitem{Kuhn1}
J.~H.~Kuhn and M.~Steinhauser,
Nucl.\ Phys.\ B {\bf 619}, 588 (2001)
[Erratum-ibid.\ B {\bf 640}, 415 (2002)]
[arXiv:hep-ph/0109084].
%
%
\bibitem{Kuhn2}
J.~H.~Kuhn and M.~Steinhauser,
JHEP {\bf 0210}, 018 (2002)
[arXiv:hep-ph/0209357].
%
%
\bibitem{Shipsey1}
see e.g. I.~Shipsey,
AIP Conf.\ Proc.\  {\bf 618}, 427 (2002)
[arXiv:hep-ex/0203033].
%
%
\bibitem{CUSB}
D.~M.~Lovelock {\it et al.},
Phys.\ Rev.\ Lett.\  {\bf 54} (1985) 377.
%
%
\bibitem{CLEO}
D.~Besson {\it et al.}  [CLEO Collaboration],
Phys.\ Rev.\ Lett.\  {\bf 54}, 381 (1985);\\
%
L.~Garren, CLEO note CBX-84-68 (October 1984).
%
%
\bibitem{Chetyrkin1}
K.~G.~Chetyrkin, J.~H.~Kuhn and M.~Steinhauser,
Nucl.\ Phys.\ B {\bf 505}, 40 (1997)
[arXiv:hep-ph/9705254].
%
%
\bibitem{Chetyrkin2}
K.~G.~Chetyrkin, A.~H.~Hoang, J.~H.~Kuhn, M.~Steinhauser and
T.~Teubner, 
Eur.\ Phys.\ J.\ C {\bf 2}, 137 (1998)
[arXiv:hep-ph/9711327].
%
%
\bibitem{Hoang1}
A.~H.~Hoang, J.~H.~Kuhn and T.~Teubner,
Nucl.\ Phys.\ B {\bf 452}, 173 (1995)
[arXiv:hep-ph/9505262].
%
%
\bibitem{Hoang2}
A.~H.~Hoang and T.~Teubner,
Nucl.\ Phys.\ B {\bf 519} (1998) 285
[arXiv:hep-ph/9707496].
%
%
\bibitem{Broadhurst1}
D.~J.~Broadhurst {\it et al.}, 
Phys.\ Lett.\ B {\bf 329}, 103 (1994)
[arXiv:hep-ph/9403274].
%
%
\bibitem{Baikov1}
P.~A.~Baikov, V.~A.~Ilyin and V.~A.~Smirnov,
Phys.\ Atom.\ Nucl.\  {\bf 56} (1993) 1527
[Yad.\ Fiz.\  {\bf 56N11} (1993) 130].
%
%
\bibitem{PDG}
K.~Hagiwara {\it et al.}  [Particle Data Group Collaboration],
Phys.\ Rev.\ D {\bf 66}, 010001 (2002).
%
%
\bibitem{ARGUS}
H.~Albrecht {\it et al.}  [ARGUS Collaboration],
Z.\ Phys.\ C {\bf 65}, 619 (1995).
%
%
\bibitem{LEP2}
The LEP Electroweak Working Group et al., 
arXiv:hep-ex/0212036.
%
%
\bibitem{Schilcher1}
J.~Bordes, J.~Penarrocha and K.~Schilcher,
arXiv:hep-ph/0212083.
%
%
\bibitem{Steinhauser1}
M.~Steinhauser, private communication.
%
%
\bibitem{Hoangcharm}
A.~H.~Hoang and A.~V.~Manohar,
Phys.\ Lett.\ B {\bf 483}, 94 (2000)
[arXiv:hep-ph/9911461];
%
A.~H.~Hoang,
arXiv:hep-ph/0008102.
%
%
\bibitem{Erler1}
J.~Erler and M.~x.~Luo,
arXiv:hep-ph/0207114.
%
%
\bibitem{Eidemuller1}
M.~Eidemuller,
arXiv:hep-ph/0207237.
%
%
\end{thebibliography}

\end{document}